**Christopher K. Allsup, Irene S. Gabashvili**
Aurametrix, USA
https://aurametrix.com

# Modeling the Dynamics of Growth in Master-Planned Communities

## Abstract

This paper describes how a time-varying Markov model was used to forecast housing development at a master-planned community during a transition from high growth to low growth. Our approach draws on detailed historical data to model the dynamics of the market participants, producing results that are entirely data-driven and free of bias. While traditional time series forecasting methods often struggle to account for non-linear regime changes in growth, our approach successfully captures the onset of buildout as well as external economic shocks, such as the 1990 and 2008–2011 recessions and the 2021 post-pandemic boom.

This research serves as a valuable tool for urban planners, homeowner associations and property stakeholders aiming to navigate the complexities of growth at master-planned communities during periods of both system stability and instability.

**Keywords**: Real Estate Economics, Econometrics, Master-Planned Communities, MPC, Housing Market Forecasting, Buildout Dynamics, Time-Varying Markov Models, Regime-Switching Models, Housing Permits, Residential Growth Patterns
Predictive Modeling, Active Adult Lifestyle Communities, Active Adult Retirement Communities, Economic Cycles

## Introduction

### Background

Master-planned communities (MPCs) have significantly evolved since their inception in the early 20th century, becoming a prominent feature of the residential landscape. The age-specific focus of MPCs gained prominence with the rise of active adult lifestyle communities (AALCs), pioneered by Youngtown, AZ and brought to widespread attention by the successful debut of Sun City on January 1, 1960. Drawing significant crowds, Sun City featured a variety of home models, a shopping center, a recreation center, and a golf course, setting a new standard for retirement living [1].

Over the past six decades, AALCs have increasingly attracted a broader range of residents, including younger families seeking to tap into the recreational, social, and safety benefits such unincorporated towns and villages provide. It is not unreasonable to imagine a future in which a growing number of these communities cater to residents on the basis of shared cultural, political, and lifestyle dimensions, potentially reshaping patterns of social organization and urban development.

Planned housing developments typically have a predetermined number of platted lots allocated for single-family homes that are sold to prospective homeowners, builders, and investors over time. Although some communities expand the pool of platted lots (a notable example being The Villages in Florida, which does so on a recurring basis [2]), most permanently limit the number of homes that can be constructed.

Forecasting housing growth in these communities is consequential for two reasons: From a financial perspective, permitting fees for the development of new homes are often a major source of revenue for the community, and predicting their revenue streams becomes integral to each year's budgeting process. From an operational standpoint, accurate population predictions give community management insight into when and how much additional capacity will be needed for recreation facilities, meeting rooms, clubhouses, parking lots, and other amenities.

The case study for this work is Tellico Village, an AALC of more than five thousand homes located in East Tennessee. As Tellico Village entered 2024, the community faced the possibility of a slowdown in permitting activity due to the approach of buildout, defined herein as the maximum percentage of lots expected to be permitted.

### Rationale

This work aims to address critical gaps in existing forecasting methods by developing a model that accounts for the dynamic and regime-switching nature of housing development in master-planned communities

Traditional forecasting methods for predicting housing growth perform adequately during steady-state periods of sustained growth, but tend to be less accurate during transitions between different growth regimes. For instance, while linear regression outperformed other models during the 2016–2020 period (based on MAE, MSE, and RMSE metrics), it struggled significantly during periods of relative instability or downtrends. Figure 1, derived from data detailed later in this paper, illustrates annual construction completions (used as a proxy for new permits) at Tellico Village. The housing recession in 2007 and 2008 had a negative impact on construction in subsequent years. Yet the linear regression forecast underestimates the decline in the years following 2008, highlighting the limitations of traditional models in capturing regime changes.

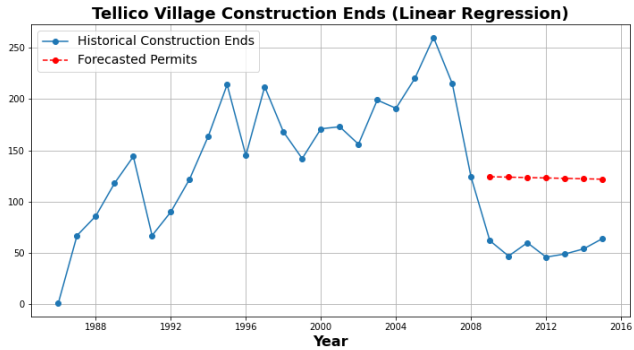

*Figure 1. Linear forecasting models have difficulty predicting transitions to new growth regimes such as the onset of a housing recession.*

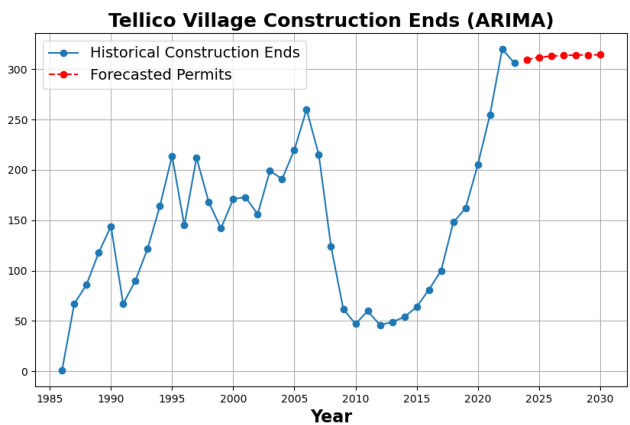

*Figure 2. ARIMA forecast does not have the information needed to predict a decline in permits due to approaching buildout.*

Another type of a regime transition that cannot be accurately predicted by linear forecasting models is the decrease in home building due to the approach of lot buildout. Buildout is the maximum percentage of lots expected to be permitted (typically not 100% because some lots might not be sellable in the foreseeable future for a variety of reasons). As of the end of 2023, Tellico Village had permitted about 83% of platted lots, yet Figure 2 indicates an autoregressive integrated moving average (ARIMA) forecast is insensitive to the approaching buildout "squeeze". And like ARIMA for yearly permit authorizations, seasonal ARIMA applied to monthly data struggled with capturing non-linear regime shifts (see Appendix 1).

The forecasting techniques highlighted here rely solely on historical permit counts and so cannot be expected to accurately forecast new growth regimes. To address the limitations of conventional forecasting techniques, a housing forecast method is needed that can model the complex buying, selling, and permitting behavior of lot owners.

## Methods and Results

### Data Acquisition and Mining

We first mined detailed historical data to understand the behaviors of the MPC market participants. The data acquisition involved the following steps:

- **Street and Lot Identification:** We extracted street names from directories and public maps within the studied community of Tellico Village.

- **Property Data Acquisition:** Data on property ownership, lot sizes, construction dates, and sales history were scraped from the Tennessee Comptroller of the Treasury's [Tennessee Property Assessment Data](#) (TPAD), covering two jurisdictions and the seven neighborhoods comprising Tellico Village.

- **Permit Data Collection:** We obtained building permit data, including applications and approvals, from the [Tellico Village Property Owner's Association](#) web pages to assess construction activities.

We next systematically integrated property assessment data with building permit applications by matching street names and lot numbers, enabling us to establish correlations between ownership and construction activities. This integration provided critical insights into development patterns within the communities.

To trace lot owner activities and understand how ownership dynamics influence development timing and outcomes, we aligned property ownership records with permit application data. By cross-referencing the names of lot owners with local contractors and permit applicants, we identified those directly involved in construction. A detailed geographic analysis further allowed us to detect lots owned by the same individual or entity, even when located on different streets but sharing common boundaries. This was essential for understanding strategic ownership decisions, such as prioritizing privacy or planning for future development.

For data from 2019 to December 2023, we utilized information from the Architectural Control Committee (ACC), the permit-authorizing body for Tellico Village. This dataset enabled us to accurately count permits and differentiate between those issued to builders and to prospective owners. We also analyzed the "Actual Year Built" from TPAD (see Figures 1 and 2), assuming permits were authorized a year prior, and examined sales data to identify houses sold shortly before or after completion to classify the lot owner as a builder. This analysis, combined with adjustments for lots sold with unfinished houses, helped us estimate the historical proportion of custom-built versus contractor-built spec homes, allowing for extrapolation of these proportions to permit authorizations.

### Modeling the Dynamics of Market Participants

The data extraction and mining methods described above provided the insight needed to categorize the market participants by their transaction behavior. These are the owner categories we uncovered:

**"Builders":** Home builders/contractors. Builders usually sell or permit undeveloped lots within 2 years of purchase.

**"Prospects":** Prospective residents that have purchased a single lot for permitting. Some Prospects might purchase a lot years in advance of development. They typically live in the newly built house for at least 3 years before selling it.

**"Adjacents":** Homeowners that own neighboring lots in addition to their lot of residence. These owners do not permit the lots and sell them only when they leave the community.

**"Flippers":** Investors that often own multiple lots and never permit. Flippers turn over lots relatively quickly.

Initially, our plan called for extracting time-to-permit distributions for each corresponding lot category, as illustrated conceptually in Figure 3, and then applying Bayesian analytics to forecast annual permits (see Appendix 2). However, lots are not locked into their respective categories and can instead undergo one or more changes in ownership prior to being permitted, as illustrated in Figure 4. This is indeed an intrinsic characteristic of Adjacent and Flipper lots, which are always sold before being eventually permitted. Accurately determining time-to-permit distributions was challenging due to the TPAD's limited one-year visibility into lot ownership. However, transaction data like pricing, legal instruments, and transaction types was available for all preceding years.

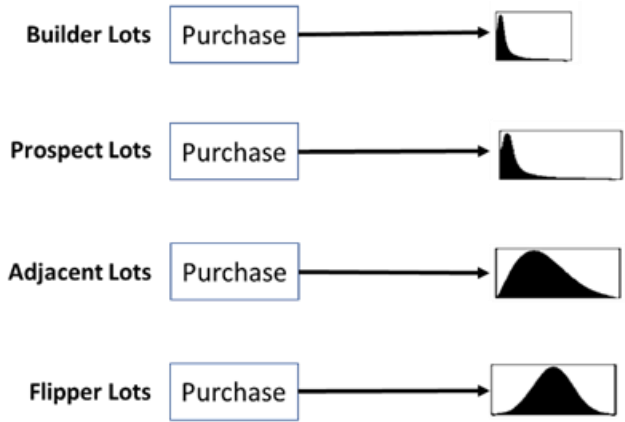

*Figure 3. Conceptual time-to-permit probability distributions for the different lot categories.*

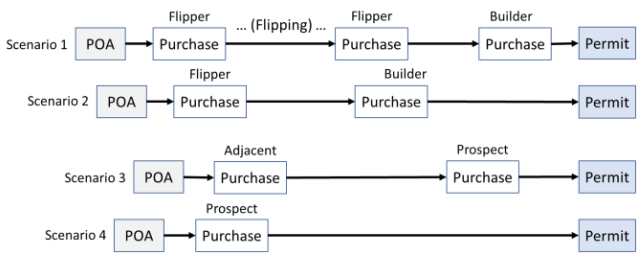

*Figure 4. A few examples of how a lot can change ownership from the time it is initially purchased from a property owners association until it is finally permitted.*

A better approach was to model the complex buying, selling, and permitting interactions among the four ownership categories as an evolving Markov process [3], as shown in the state diagram of Figure 5.

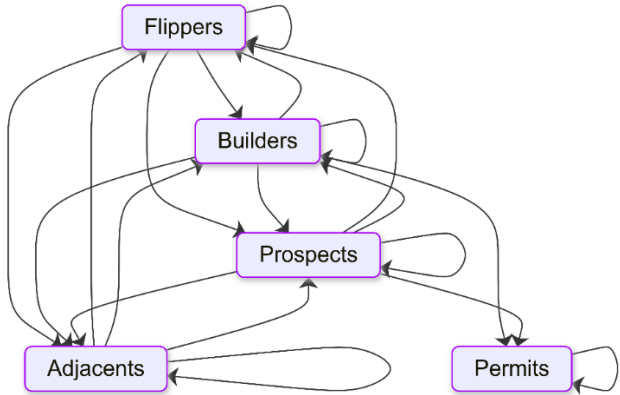

*Figure 5. Unpermitted lots categorized by owner behavior evolve over time as a Markov process.*

The blocks represent the number of unpermitted lots in each owner category. During any discrete time interval t, a certain percentage of lots of a given type are sold to another type (represented by arrows flowing out). All possible combinations of buying, selling, and holding are accounted for except that only Builders and Prospects lots can be permitted. The Permits block itself represents the number of cumulative permits.

The evolution of permitted and unpermitted lots behaves like a Markov process comprised of states that evolve over discrete time intervals. The next state $x_{t+1}$ is always a linear function of the current state $x_t$:

(1) $x_{t+1} = x_t P_t$

where $x_t = [\text{Flippers}, \text{Builders}, \text{Prospects}, \text{Adjacents}, \text{Permits}]$ and $P_t$ is the state transition matrix:

| t+1: | Flippers | Builders | Prospects | Adjacents | Permits (R) |
|---|---|---|---|---|---|
| t: Flippers | Pr{F→F} | Pr{F→B} | Pr{F→P} | Pr{F→A} | 0 |
| t: Builders | Pr{B→F} | Pr{B→B} | Pr{B→P} | Pr{B→A} | Pr{B→R} |
| t: Prospects | Pr{P→F} | Pr{P→B} | Pr{P→P} | Pr{P→A} | Pr{P→R} |
| t: Adjacents | Pr{A→F} | Pr{A→B} | Pr{A→P} | Pr{A→A} | 0 |
| t: Permits (R) | 0 | 0 | 0 | 0 | 1 |

The elements in the matrix predict the next year's state values. For historical events, they are observed percentages; for future events, they are probabilities. Notice all rows must sum to 100%, the diagonal elements reflecting the fact that each lot category retains all the lots not sold or permitted by its owners. Therefore, for each of the 4 unpermitted states, one transition probability can be determined by the other 3 (4 for Builders and Prospects). This implies there are 5 equations in 14 unknowns, and the solution space for the transition probabilities is infinite. Moreover, the transition matrix itself can change from one time step to the next.

To arrive at a viable solution for each state we used constraints based on historical data, as described below.

### Dynamics of Ownership Transitions and Regime Shifts

Our analysis employed multiple constraint scenarios to capture temporal variability and regime shifts in the transition dynamics. Core constraints applied across all scenarios included the total number of platted lots, the number of permits authorized from 1985 to 2023 (approximated from the actual year the house was built before 2019), and the evolving ratio between contractor-built (spec) homes and custom-built homes. Similar to the U.S. national data [4], there was a clear trend toward buyers' favoring spec homes over custom-built homes, and the ratio of custom-built homes was always increasing during economic downturns and decreasing during boom cycles (see Appendix 3). This ratio is a critical factor in understanding the dynamics between the Prospects and Builders groups. Each scenario was designed to align the range of lot ownership categories—Flippers, Builders, Prospects, and Adjacents—with observed data for each year, based on sales transactions and years of construction end.

Initially, using sales data, we estimated time-varying transition probabilities by calculating local one-hop transitions between states, providing short-term predictions of state changes. This was followed by an iterative process that refined these models through optimal decision-making, akin to linear programming. This approach allowed us to optimize decisions within finite state and action spaces while accounting for temporal variations.

Transition probabilities were modeled based on the observed dynamics of land buying, selling, holding, and permitting among different groups. Each scenario was tailored to provide realistic estimates while maintaining flexibility in key transitions, such as Flippers-to-Builders, Flippers-to-Prospects, and Builders-to-Permits. To address the challenges of over-constrained periods like 1990, 2008, and 2021, we adjusted the scenarios to accommodate larger variations in permitting behavior, enabling the model to fit during periods of economic and market fluctuations.

The fitting procedures revealed significant year-over-year variations in transition probabilities, indicating regime shifts across multiple periods. The Lagrange Multiplier (LM) test confirmed statistically significant regime switching in key transitions. Clustering techniques, including k-means and Gaussian Mixture Models (GMM), identified distinct periods of transition behavior, with major shifts in 1990, 2005, 2008, and 2021, suggesting these shifts were influenced by external factors.

To further analyze regime shifts, CUSUM charts were used to track the evolution of transitions, revealing significant changes during economic downturns and potential slowdowns after 2023, consistent with regime changes due to buildout. Additionally, Information Criteria, specifically AIC and BIC, were employed to determine the optimal number of regimes. While AIC suggested that models with more regimes fit the data better, BIC indicated that a 3-regime model offered the best balance between model fit and complexity.

This comprehensive analysis and determination of transition probabilities formed the foundation of a robust, time-varying Markov model (TVMM) that more accurately reflects the complex, time-dependent evolution of lot ownership and development within master-planned communities.

### Forecasting Transition Probabilities

Once recent historical transition probabilities have been estimated, we can employ exponential moving average (EPA) to capture any trend in the data that would signal how the probabilities might evolve over future cycles. EPA recalculates the current value in the historical data as a smoothed value function of $\alpha$:

(2)  smoothed_value = $\alpha$ · current_value + $(1 - \alpha)$ · previous_smoothed_value

The value of $\alpha$ must be consistent with the time evolution of lot categories. After the states were forecast (see next section), it was observed that adjacent lots were steadily increasing over time. An $\alpha$ of 0.15 was then selected to place more weight on the previous smoothed value, thereby ensuring adjacent lots would steadily decrease over time.

When calculating future smoothed values, current_value is replaced by a random variable that deviates a small amount from the previous smoothed value to avoid reproducing the same probabilities every cycle:

(3)  smoothed_value = $\alpha$ · random_value + $(1 - \alpha)$ · previous_smoothed_value

in which random_value value is a random variable with mean = previous_smoothed_value and standard deviation = 0.05.

After the future smoothed values have been calculated, a three-step normalization procedure is applied to ensure 1) the outbound probabilities for each state sum to 100%, 2) that all values are positive by clipping negative values to 0, and 3) that the probabilities sum to 100% after clipping.

It is possible to model future high- and low-growth scenarios by applying a scaling factor to the last historical smoothed value, effectively altering the mean values of the forecasted probabilities. For example, to model a future recession one can decrease the rate of future permitting by decreasing the scaling factor and increasing the standard deviation of random_value.

### Adaptive Forecasting for Permits

The forecasted transition probabilities were loaded into the Markov model along with the current year's estimated number of lots in each category: state values for Flippers, Builders, Prospects, Adjacents, and Permits. The states were then updated using (1) each year across the time horizon of the forecast while keeping track of the number of annual permits issued.

Figure 6 provides insight into determining how many historical cycles are needed to detect a trend. The chart displays estimated historical and future lots that remain unpermitted in each owner category. Both the number of Builders and Prospects – the most important indicators of housing growth – begin to decline in 2021 and decline each year through the last historical year, 2023. The data to detect these declines is contained in the transition probabilities needed to predict the states in 2020 through 2023. We would therefore conclude from these estimates that not more than four years of historic data are needed to detect the downward trends.

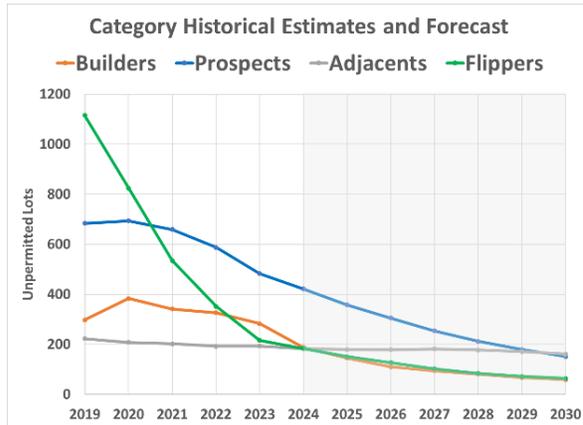

*Figure 6. Analysis of unpermitted categories demonstrates the need for only four years of historical estimates of transition probabilities for the Tellico Village forecast.*

Monte Carlo simulations were run on the forecast to calculate 95% confidence intervals based on a 20% standard deviation of all state transition probabilities. Figure 7 displays the results for both annual permits and buildout percentage (single-family permitted lots ÷ platted lots). The forecast predicts 93% buildout by the end of the decade. As mentioned in the previous section on forecasting transition probabilities, low- and high-growth forecasts can be generated for purposes of scenario planning by making blanket adjustments to the probability means and standard deviation of the random noise added during EPA.

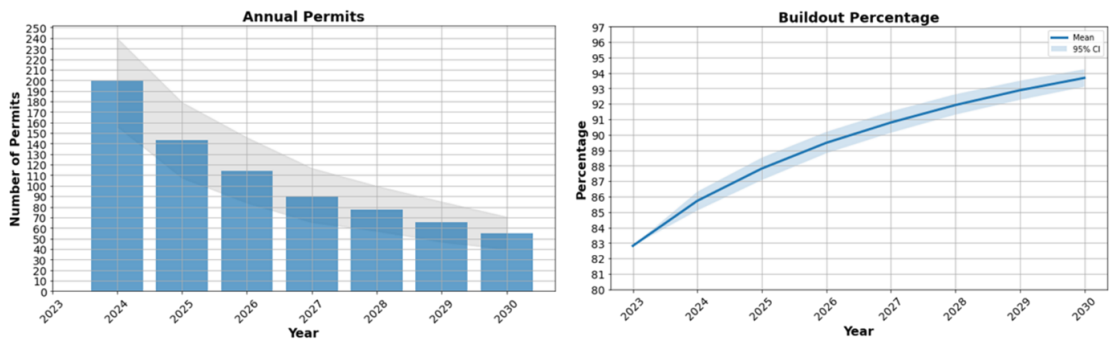

*Figure 7. The Markov model predicted a 29% decline in permits in 2024 at Tellico Village, and even slower growth ahead due to the onset of buildout.*

## Discussion

### Key Insights

This study highlights the value of using time-varying Markov models to forecast housing development within master-planned communities. By employing such a model, we were able to capture the dynamic nature of lot ownership changes and permitting behaviors over several decades, and thereby account for different growth regimes such as those that occurred during the 1990 recession, the housing crisis of the late 2000s, and the post-pandemic construction surge in 2021.

As discussed in the "Dynamics of Ownership Transitions and Regime Shifts" section, accurately estimating transition probabilities from ongoing sales data was crucial for forecasting permit authorizations with precision. Sales data provided us with insights into the likely transitions between different ownership states. For example, when a lot was purchased by the homeowner of an adjacent lot or by a known local contractor, we could infer specific future actions, such as the likelihood of immediate development or prolonged holding. These transitions, represented as probabilities, are not static; they shift in response to different growth regimes.

This dynamic nature of transition probabilities acts as an early indicator of changes in market conditions. For instance, in a period of economic expansion, a higher probability of lots transitioning to builder ownership might signal an impending surge in permit applications and construction activity. Conversely, in a regime characterized by economic downturns or nearing buildout, the transition probabilities might shift toward longer holding periods or reduced development activity, reflecting a more cautious market stance.

By integrating these probabilistic transitions into our Markov model, we gained the ability to forecast not just based on past trends but also on the evolving behaviors of market participants. This approach allows for more responsive and accurate predictions, particularly during periods of regime shifts, where traditional linear models may fail to capture the nuances of market dynamics. The early identification of these shifts through transition probabilities enhances the model's forecasting capabilities, enabling stakeholders to anticipate changes and adjust strategies accordingly. To further broaden the model's robustness, we employed Monte Carlo simulations, which accounted for uncertainties in transition probabilities and provided confidence intervals for our forecasts.

Our model demonstrated adaptability to changing market conditions essential for sustainable long-term development. One insight gained from data mining is the shift in buyer preferences during different economic cycles. During periods of economic growth, there is a noticeable trend toward quicker, ready-to-move-in options, with buyers increasingly favoring model/spec (contractor-built) homes over custom-built ones (see Appendix 3). In contrast, during times of economic uncertainty, the market is primarily driven by prospective residents who prefer to custom-build their homes, reflecting a more cautious and personalized approach to homeownership. This differentiation in buyer behavior highlights the importance of tailoring development strategies to the prevailing economic environment, ensuring that community planning remains responsive to the evolving needs of the market.

Clustering techniques, such as k-means and Gaussian Mixture Models, were employed to confirmed the presence of changes in growth regimes, further validating the importance of incorporating such transitions into predictive models. The clustering results revealed three primary regimes in the community's development:

- **Stable Growth:** Characterized by consistent activity in land buying, selling, and permitting, this regime represents periods of steady economic conditions with predictable housing market dynamics.

- **Expansion and Adjustment:** Marked by heightened market activity, this regime is typical of economic recovery or boom periods, where rapid development occurs but is eventually tempered by market adjustments, such as changes in interest rates or regulatory policies.

- **Constrained Activity:** Reflective of times when economic or policy constraints limit market activity, leading to sharp declines in certain transitions and less predictable permitting behavior. The buildout regime also constrains housing growth, primarily due to the scarcity of available lots and increased resistance from current lot holders.

These insights emphasize the importance of incorporating dynamic, data-driven approaches into forecasting models, particularly in complex systems like master-planned communities, where traditional models may fail to account for non-linear regime shifts.

## Our Contributions vs. Prior Work

Our research advances the limited and largely undeveloped field of master-planned community (MPC) growth dynamics by introducing a novel, data-driven approach to modeling transitions and regime shifts, particularly as communities approach buildout.

While existing studies have examined broader aspects of community development and the potential of MPCs for innovative health and social research [5,6], they either overlook the critical analysis of housing construction dynamics or relied on traditional forecasting methods. The regression model of Smersh, Smith, and Schwartz [7], for example, while identifying important determinants of housing development, could not explain a significant portion of the growth pattern.

By employing advanced statistical techniques, including time-varying Markov models and clustering methods, our research provides insights into the regime shifts and behavioral changes that occur as MPCs evolve.

Regime-switching Markov models, pioneered by Hamilton for analyzing U.S. GDP growth [3], have influenced numerous fields beyond economic cycles. Later methodological advancements, such as those by Bazzi et al. [8], underscored the potential of time-varying regime-switching Markov models, where transition probabilities dynamically adapt based on predictive likelihood functions. This adaptation offers a more nuanced and responsive method for modeling complex systems. However, the field still faces significant challenges, particularly the absence of robust, off-the-shelf optimization strategies, which drives the need for continued development of more sophisticated models. These advancements include modifications such as second-order time-invariant Markov transition probabilities [9] and higher-order models capable of capturing both short-term dependencies and long-range patterns.

Our study builds on these advancements by focusing on the micro-level interactions of different lot ownership categories within an MPC. Leveraging a time-varying Markov Model, we developed a framework that models the transitions between these owner categories, which are crucial to understanding the evolution of lot permitting and construction activity.

Utilizing actual permit data, we addressed the complexities that arise in the final stages of MPC development, where constraints such as limited lot availability and resistance from current lot holders can slow the pace of construction. Studies like those by Shakro [10] from Tufts University have highlighted the importance of detailed permit data in understanding development patterns, but they did not extend this analysis to the micro-level interactions within MPCs.

As in this work, Adamec [11] used advanced stochastic modeling techniques, applying a two-state Markov-Switching model to analyze construction sector performance. While Adamec's work emphasized the stochastic nature of construction time series, it was based on only two states and focused only on external environmental influences such as the weather and regulatory factors such as tax rate.

Our study diverges from [11] by narrowing the focus to micro-level interactions and decision-making within MPCs, providing a detailed, behavior-driven perspective that more accurately reflects the realities of these unique communities. This granular approach not only enriches the understanding of MPC development but also introduces the potential for more precise predictions and strategic planning.

## Conclusions

This study is the first to analyze large volumes of state government and local MPC authority records for a comprehensive analysis of housing development patterns, demonstrating the potential of this data to train next-generation forecasting models. By employing a time-varying Markov model, we captured the complex interactions between ownership categories and their impact on permitting behaviors, particularly during transition dynamics and regime shifts. Our approach effectively accounted for non-linear changes, such as those caused by approaching buildout and economic shocks.

Our work highlights the importance of incorporating dynamic, data-driven approaches to forecasting in master-planned communities. This approach not only improves the accuracy of predictions during periods of uncertainty but also lays the groundwork for advanced AI-driven models that can further enhance our understanding of housing development patterns.

As communities like Tellico Village near buildout, these insights will be invaluable for urban planners, developers, property owner associations, homeowner associations, and policymakers in managing sustainable growth.

## Acknowledgements

We appreciate the coding support provided by Claude Sonnet 3.5 and ChatGPT 4o.

## Funding

This research received no external funding.

## Data Availability Statement

This study was conducted as an open-source collaborative project through the [Open Science Framework](#), registered under DOI [10.17605/OSF.IO/TGV6Q](#). Associated code is available at [GitHub](#): https://github.com/MPC-Dynamic/Growth

## Conflicts of Interest

None declared.

## Appendix 1: SARIMA Analysis

The SARIMA(1,1,1)×(1,1,1,12) model was applied to monthly permit data from 2010 to 2024, capturing both short-term fluctuations and yearly seasonal patterns. The autoregressive (AR) term is near zero, indicating minimal influence from recent past values on future data. However, the moving average (MA) term is significantly negative, showing that previous shocks are smoothed out over time. The accompanying plot illustrates the observed data, smoothed values from Holt-Winters exponential smoothing, and the SARIMA forecast extending through 2028. The forecast suggests that historical seasonal trends will continue, with periodic fluctuations. However, since the SARIMA model is linear, it cannot account for the expected shift to a restricted growth regime in the future.

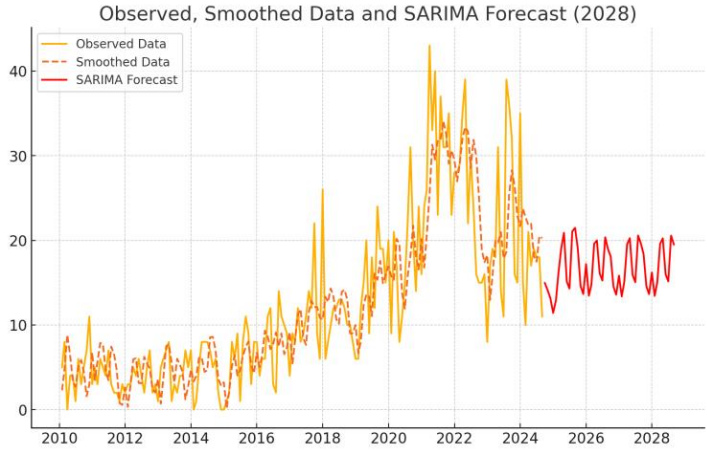

*Supplemental Figure 1: Monthly permit data, exponentially smoothed monthly data, and SARIMA forecast. Linear forecasting models have difficulty predicting transitions to new growth regimes such as the buildout regime.*

## Appendix 2: Estimating Expected Annual Lot Permits Using Time-to-Permit Probability Distributions

The expected number of lots permitted each year can be estimated the following way: Let $P_c(t)$ be the discrete probability distribution of time in years from purchase to permit of lot category c based on historical data. If an unpermitted lot i belonging to the category was purchased $t_i$ years ago, then $L_c(t)$, the expected number of lots in category c permitted in year t, is the sum of the posterior probability distributions for all lots in the category:

$$(4)\quad L_c(t) = \sum_{i=1}^{N} P_c(t+t_i \mid t_i) \quad t \geq 0$$

where N is the number of unpermitted lots in the category. The total expected number of lots permitted each year, L(t), is the sum of the expected number of permitted lots across all 4 categories:

$$(5)\quad L(t) = \sum_{c=1}^{4} L_c(t) \quad t \geq 0$$

Given the time-to-permit probability distribution $P_c(t)$ of lot category c, we can use Bayes' Theorem to calculate the posterior probability distribution for an individual lot i that has been purchased a number of years $t_i$ in the past:

$$(6)\quad L_{ci}(t) = P_c(\text{permitted after } t + t_i \mid \text{not permitted up to } t_i)$$

$$= \frac{P_c(\text{not permitted up to } t_i \mid \text{permitted after } t + t_i) \cdot P_c(\text{permitted after } t + t_i)}{P_c(\text{not permitted up to } t_i)}$$

The first probability in the numerator is the *likelihood* term, the probability that the lot has not been permitted by $t_i$ years given that it will be permitted after $t + t_i$ years, which is 100%.

The second probability in the numerator is the *prior* term, the probability that the lot is permitted after $t + t_i$ years, which is just the probability $P_c(t + t_i)$.

The probability in the denominator is the *marginal likelihood* term, the probability that the lot is not permitted by $t_i$ years, which is the same as the probability that the lot is permitted after $t_i$ years:

$$(7) \quad P_c(\text{not permitted up to } t_i) = \sum_{k=t_i+1}^{\infty} P_c(k)$$

Dividing the numerator in (6) by the quantity calculated in (7) normalizes the distribution. The expression for the posterior probability distribution for a given lot i reduces to:

$$(8) \quad L_{ci}(t) = P_c(t + t_i) \div \sum_{k=t_i+1}^{\infty} P_c(k)$$

### Appendix 3: Estimating Custom- vs. Contractor-Built Developments

For the period from 2019 to 2024, our estimates of custom-built versus contractor-built developments are accurate due to detailed records of permit ownership and comprehensive contractor lists. However, estimates for earlier years are less certain, primarily because of incomplete data on construction completion dates and missing transactions in the database. We infer that a house was contractor-built if the lot was sold shortly before or after construction was completed. U.S. data were sourced from the Survey of Construction (SOC), a study conducted jointly by the Department of Housing and Urban Development (HUD) and the U.S. Census Bureau [3]. This survey is based on a multi-stage stratified random sample, providing a broad overview of national trends.

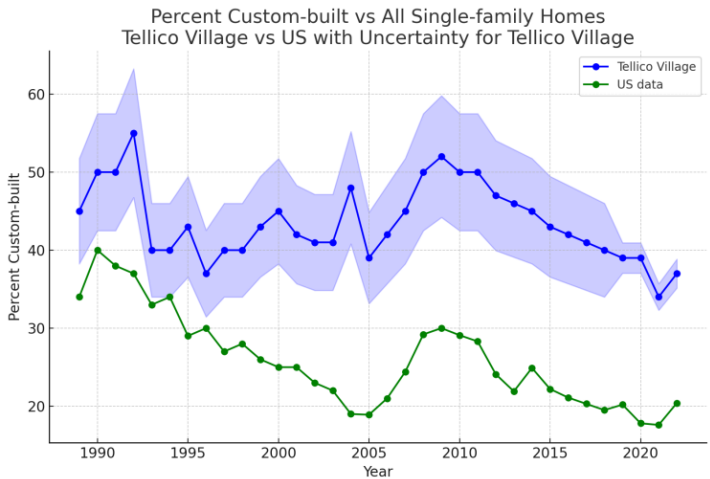

*Supplemental Figure 2: Ratio of custom-built houses to all single-family houses in Tellico Village (blue) compared to the U.S. average (green). The Tellico Village data includes a 95% confidence interval, with a standard deviation of approximately 15% of the values. The uncertainty significantly decreases in the years 2019-2024 due to the availability of more precise data.*

The Pearson correlation between the U.S. data and the Tellico Village data is approximately 0.553, with a p-value of 0.0007, indicating a statistically significant moderate positive correlation.

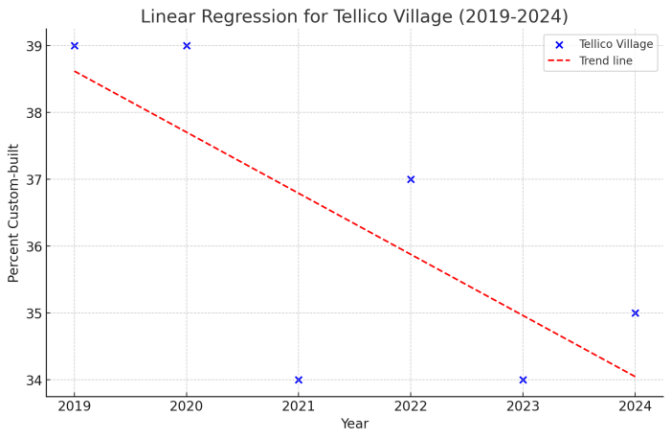

*Supplemental Figure 3: Ratio of custom-built houses to all single-family houses in Tellico Village during 2019-2024 according to permit authorization data. 2024 is incomplete, covering only January to August.*

Linear regression analysis for permit authorizations in Tellico Village in 2019-2024 this shows a strong negative correlation with an R-value of approximately -0.732. The p-value of 0.098 indicates moderate statistical significance, though it is not below the conventional threshold of 0.05. The 95% confidence interval for the slope of the linear regression from 2019 to 2024 ranges from -2.10 to 0.27. The wide interval suggests a downward trend, but the slope is not tightly constrained, indicating uncertainty in the exact magnitude of this trend. Linear regression analysis across the entire dataset also yields a moderately significant slight negative trend over time. US data suggest a more significant trend, but since Tellico Village data are pertaining to specific population, the trends are expected to not completely align and even be divergent in some time intervals.

**Abbreviations**

AALC: Active Adult Lifestyle Community
ACC: Architectural Control Committee
AIC: Akaike Information Criterion
ARIMA: Autoregressive Integrated Moving Average
BIC: Bayesian Information Criterion
CUSUM: Cumulative Sum
EMA: Exponential Moving Average
GMM: Gaussian Mixture Model
HUD: Department of Housing and Urban Development
LM: Lagrange Multiplier

MAE: Mean Absolute Error
MCMC: Markov Chain Monte Carlo Method
MPC: Master-Planned Community
MSE: Mean Squared Error
OSF: Open Science Framework
POA: Property Owners Association
RMSE: Root Mean Square Error
SARIMA: Seasonal Autoregressive Integrated Moving Average
SOC: Survey of Construction
TPAD: Tennessee Property Assessment Data
TVMM: Time-Varying Markov Model